\documentclass[]{iopart}

\expandafter\let\csname equation*\endcsname\relax
\expandafter\let\csname endequation*\endcsname\relax
\usepackage[perpage,marginal]{footmisc}
\usepackage[unicode]{hyperref}
\usepackage[all]{hypcap}
\usepackage{amssymb,amsmath,verbatim,mathtools,needspace,enumitem,etoolbox,graphicx,microtype}
\usepackage[usenames,dvipsnames]{xcolor}
\usepackage[T1]{fontenc}
\usepackage[utf8]{inputenc}
\usepackage{pifont}
\usepackage{yfonts}
\usepackage{url}
\usepackage[normalem]{ulem}
\definecolor{linkcolor}{rgb}{0.0,0.3,0.5}
\hypersetup{colorlinks=true,linkcolor=linkcolor,citecolor=linkcolor,filecolor=linkcolor,urlcolor=linkcolor}

\newcommand{\caltech}{{TAPIR 350-17, California Institute of Technology, 1200 E California Boulevard, Pasadena, CA 91125, USA}}
\newcommand{\rochester}{{Center for Computational Relativity and Gravitation, Rochester Institute of Technology,
85 Lomb Memorial Drive, Rochester, NY 14623, USA}}
\newcommand{\olemiss}{{Department of Physics and Astronomy, The University of Mississippi, University, MS 38677, USA}}
\newcommand{\jhu}{{Department of Physics and Astronomy, Johns Hopkins University, 3400 N. Charles
Street, Baltimore, MD 21218, USA}}
\newcommand{\dallas}{{Department of Physics, The University of Texas at Dallas, Richardson, TX 75080, USA}}
\newcommand{\bowdoin}{{Department of Physics and Astronomy, Bowdoin College, Brunswick, ME 04011, USA}}
\newcommand{\cambridge}{{Department of Applied Mathematics and Theoretical Physics, Centre for Mathematical Sciences, University of Cambridge, Wilberforce Road, Cambridge CB3 0WA, UK}}
\newcommand{\bham}{School of Physics and Astronomy and Institute for Gravitational Wave Astronomy, University of Birmingham, Birmingham, B15 2TT, UK}

\begin{document}

\begin{center}
\title[D.~Gerosa et al.]{{Wide nutation}:
binary black-hole spins repeatedly oscillating  from full alignment to full anti-alignment}
\end{center}

\author{
Davide~Gerosa$^{*\,1,2}$,
Alicia~Lima$^{1,3}$,
Emanuele~Berti$^{4, 5}$,
Ulrich~Sperhake$^{6,1,5}$,
Michael Kesden$^{7}$, \\
Richard O'Shaughnessy$^{8}$
}
\address{$^{1}$~\caltech}
\address{$^{2}$~\bham}
\address{$^{3}$~\bowdoin}
\address{$^{4}$~\jhu}
\address{$^{5}$~\olemiss}
\address{$^{6}$~\cambridge}
\address{$^{7}$~\dallas}
\address{$^{8}$~\rochester}

\address{*~Einstein Fellow}

\ead{\href{mailto:dgerosa@caltech.edu}{dgerosa@caltech.edu}}

\begin{abstract}
Within the framework of 2PN black-hole binary spin precession, we explore configurations where one of the two spins oscillates from being completely aligned with the orbital angular momentum to being completely anti-aligned with it during a single precession cycle. This \emph{{wide nutation}} is the extreme limit of the generic phenomenon of spin nutation in black-hole binaries. Crucially, {wide nutation} happens on the short precession time scale and it is not a secular effect due to gravitational-wave radiation reaction. The spins of these binaries, therefore, flip repeatedly as one of these special configurations is entered. %
Binaries with total mass $M$, mass ratio $q$, and dimensionless spin $\chi_1$ ($\chi_2$) of the more (less) massive black hole are allowed to undergo %
{wide nutation} at binary separations $r \leq r_{\rm wide} \equiv [(q \chi_2 - \chi_1)/(1-q)]^2 M$. %
Sources that are more likely to {nutate widely} have similar masses and effective spins close to zero.
\end{abstract}

\section{Introduction}

The detection of gravitational waves (GWs) from compact binary inspirals offers the unprecedented opportunity to observe spin precession in black hole (BH) binaries.
Measurements of high-order spin-spin and spin-orbit couplings may be challenging~\cite{2014PhRvL.112y1101V,2016PhRvD..93d4071T,2017PhRvD..96l4041W}, but they will prove crucial to solving some of the most outstanding problems in GW astronomy. Orbital-plane and spin precession have been shown to break degeneracies in the detected signals, thus allowing us to better characterize the growing population of detected BH mergers~\cite{1994PhRvD..49.2658C,2015ApJ...798L..17C}. Measurements of spin orientations are also vital to constraining BH binary formation mechanisms: they can clarify whether stellar binaries or dynamical interactions are responsible for the majority of the event rate \cite{2016ApJ...832L...2R,2017Natur.548..426F,2017PhRvD..95l4046G} and allow us to constrain poorly understood astrophysical processes in the lives of massive stars, like fallback supernovae \cite{2018ApJ...862L...3S}, natal kicks \cite{2017PhRvL.119a1101O,2018PhRvD..97d3014W}, core-envelope interactions \cite{2017arXiv170607053B,2019MNRAS.483.3288P}, and tidal forces \cite{2013PhRvD..87j4028G,2018PhRvD..98h4036G}.

Isolated BHs in general relativity are simple systems, but the phenomenology of spinning BH binaries is a very complex and fascinating area of research. The seminal work by Apostolatos et al.~\cite{1994PhRvD..49.6274A} first highlighted most of the essential features of precessional dynamics. The orbital angular momentum $\mathbf{L}$ and the BH spins $\mathbf{S_1}$ and $\mathbf{S_2}$ all precess about the total angular momentum $\mathbf{J}$, whose direction is roughly constant during the inspiral. Unlike the motion of a spinning top, however, these three vectors do not simply move around on fixed cones: relativistic spin-orbit and spin-spin couplings give rise to complex and sometimes counterintuitive dynamics.

As already noticed by Apostolatos et al.~\cite{1994PhRvD..49.6274A}, the simplest kind of precessional dynamics, where all three vectors $\mathbf{L}$, $\mathbf{S_1}$ and $\mathbf{S_2}$ precess about $\mathbf{J}$, must inevitably be modified when the magnitude of $\mathbf{J}$ shrinks to zero and the binary loses its gyroscopic bearings. In the absence of a fixed direction anchoring the dynamics, the momenta ``tumble'' in what is now commonly referred to as ``transitional precession.'' 
A decade later, Schnittman \cite{2004PhRvD..70l4020S} identified special configurations that he called ``spin-orbit resonances'' where all three vectors are locked into a single plane and precess jointly. Soon after, the first numerical relativity simulations of BH binaries revealed that precessing configurations are responsible for the largest BH recoils, with potentially dramatic astrophysical implications~
\cite{2007PhRvL..98w1102C,2008PhRvD..77l4047B,2010ApJ...715.1006K,2011PhRvL.107w1102L,2018PhRvD..97j4049G}.
 Recent work by Lousto et al. \cite{2015PhRvL.114n1101L,2016PhRvD..93d4031L} pointed out the possibility of large nutational oscillations, which they call
``flip flops''.

 A particularly convenient framework to isolate and model BH binary spin precession arises from considering the different timescales
involved in the problem and identifying  conserved   quantities.
More specifically,
the BHs in a binary orbit about each other with period $t_{\rm orb}\propto r^{3/2}$ (where $r$ is the binary separation), emitting GWs on the radiation reaction time scale $t_{\rm rad}\propto r^4$~\cite{1964PhRv..136.1224P}. Spin precession deeply affects the binary dynamics by introducing an additional evolutionary time scale: the precessional time scale $t_{\rm pre}\propto r^{5/2}$~\cite{1994PhRvD..49.6274A}. In the post-Newtonian (PN) regime, one thus obtains the timescale hierarchy $t_{\rm orb}\ll t_{\rm pre} \ll t_{\rm rad}$ and can therefore leverage multi-timescale methods. This insight allowed further progress: the complete set of 2PN spin precession equations (see e.g.~\cite{2008PhRvD..78d4021R}) can be solved analytically on $t_{\rm pre}$ using effective potential techniques~\cite{2015PhRvL.114h1103K,2015PhRvD..92f4016G}. These solutions have been used to simplify waveform calculations~\cite{2017PhRvD..95j4004C,2017PhRvL.118e1101C}, and now serve as the backbone of state-of-the-art templates for GW data analysis~\cite{2018arXiv180910113K}.
Within the solutions found in Ref.~\cite{2015PhRvL.114h1103K,2015PhRvD..92f4016G}, spin-orbit resonances can be reinterpreted as the ``zero-amplitude'' limit of more generic {librating} configurations~\cite{2015PhRvD..92f4016G}, and the phenomenon of transitional precession can be understood as a special case of more generic nutational resonances~\cite{2017PhRvD..96b4007Z}. Effective potential techniques were also used to reveal that certain aligned-binary configurations are unstable~\cite{2015PhRvL.115n1102G,2016PhRvD..93l4074L} and to investigate discontinuous limits of the precessional dynamics~\cite{2017CQGra..34f4004G}.

In this paper we further extend these investigations. We present a new precessional phenomenon, that we call ``{wide nutation}.'' In these special configurations, one of the two spins performs a full nutation, evolving from complete alignment with $\mathbf{L}$ to complete anti-alignment {\em within a single precession cycle}. These are, by definition, the largest possible oscillations of BH spins in binary systems.

Let us consider a BH binary with mass ratio $q=m_2/m_1\leq1$, total mass $M=m_1+m_2$, dimensionless spins $0\leq \chi_i\leq 1$ %
and orbital separation $r$ 
(hereafter $G=c=1$). First, we highlight this extreme type of spin {precession} can be realized, i.e.~there exist spinning BH binary configurations such that within a precession cycle one BH's spin nutates from complete alignment with the orbital angular momentum $\mathbf{L}$ to complete antialignment with $\mathbf{L}$. A necessary condition, derived with no approximation
beyond the PN formalism, for this tumbling motion to occur is that the orbital separation satisfies 
\begin{align}
r\leq {r_{\rm wide}} \equiv \left(\frac{q \chi_2 -\chi_1}{1-q} \right)^2 M\,.
\label{rwide_intro}
\end{align}
Among the two BHs, only the one with lower dimensionless spin is allowed to undergo {wide nutations}. The more massive BH (with mass $m_1$) can {nutate widely} if $\chi_1\leq \chi_2$, while the secondary BH (with mass $m_2$) can {nutate widely} if $\chi_2\leq \chi_1$. Even though complete pole-to-pole nutation constitutes a fine-tuned condition, we find that large nutations are still possible for a large fraction of binaries at some stage during their inspiral lifetime. In particular,  {wide nutation} can take place in the sensitivity window of ground- and space-based GW detectors.

\begin{figure*}[t!]
\centering
\includegraphics[width=\textwidth]{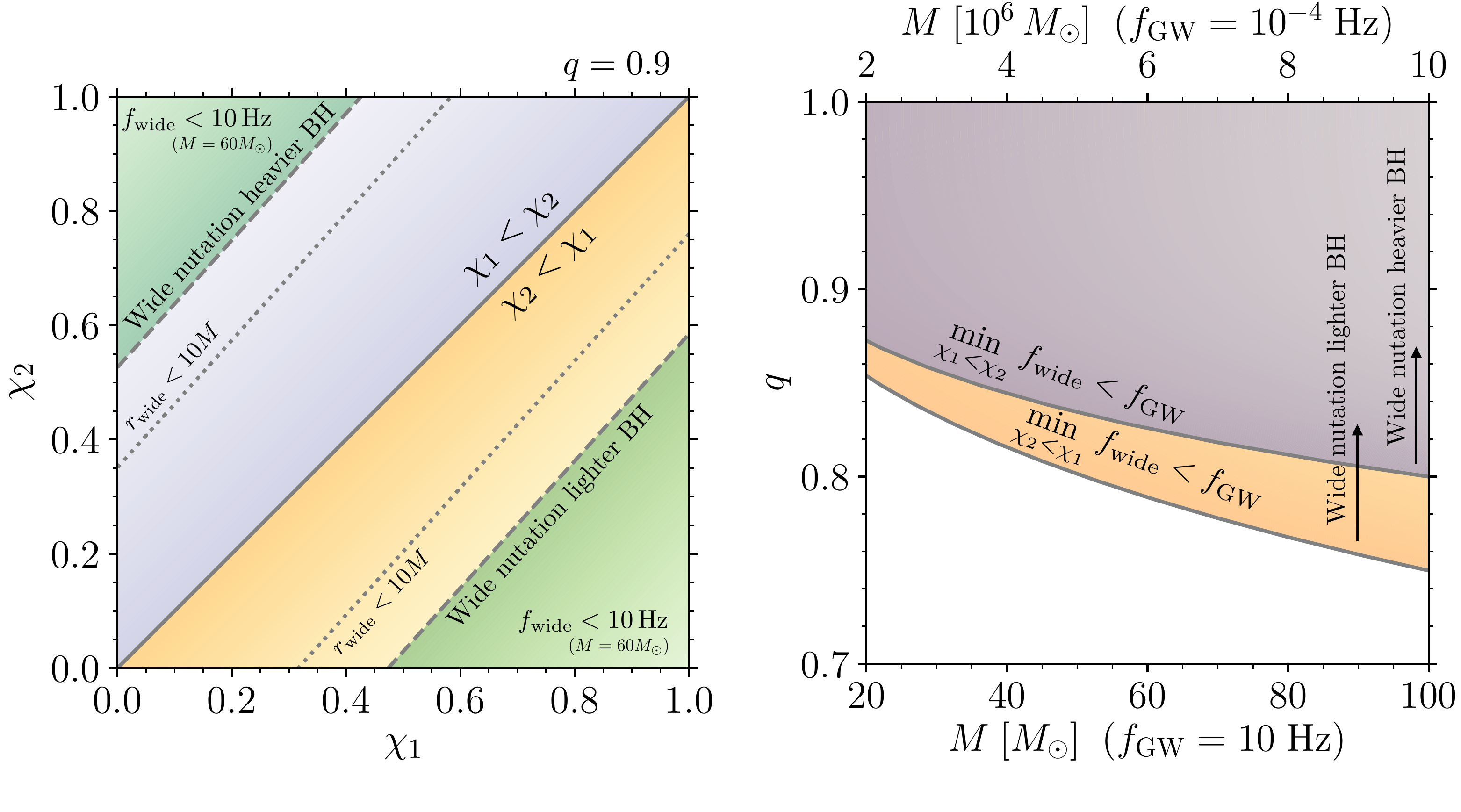}
\caption{Parameter space regions where BH binaries can {undergo wide nutation} while being observed by current and future GW detectors. {\it Left panel}: values of the spin magnitudes $\chi_1$ and $\chi_2$ that allow for {wide nutation} for fixed values of $q=0.9$ and $M=60M_\odot$. {Wide nutation} of the {heavier (lighter)} BH is possible in the region above (below) the solid line where $\chi_1<\chi_2$ ($\chi_2<\chi_1$). {Wide  nutation} can happen in band if the threshold ${r_{\rm wide}}$ is crossed before binaries become visible. For LIGO/Virgo, this is corresponds to ${f_{\rm wide}}<f_{\rm GW}= 10\,{\rm Hz}$, as shown in the two green regions marked by dashed lines. For any value of $M$ one must impose ${r_{\rm wide}}\gtrsim10M$ (dotted lines) for {wide nutation} to happen in the PN regime and be physically relevant. {\it Right panel}: values of mass ratio $q$ and total mass $M$ that allow for {wide nutation}, extremizing over the spin magnitudes. {Wide nutation} of the {heavier (lighter)} BH is possible above 
the upper (lower) solid curve.
The top and bottom x-axes are scaled with values relevant to LIGO/Virgo ($M\sim 10 M_\odot$ and $f_{\rm GW}= 10\,{\rm Hz}$) and LISA ($M\sim 10^6 M_\odot$ and $f_{\rm GW}= 10^{-4}\,{\rm Hz}$), respectively.
}
\label{wideconditions}
\end{figure*}

Figure~\ref{wideconditions} illustrates the regions of the parameter space where {wide nutation} can occur in the LIGO/Virgo or LISA band. In the left panel, we notice that  {wide nutation} is more likely to happen if the two spin magnitudes are different from each other, because the threshold ${r_{\rm wide}}$ is crossed earlier (typically before entering the detector's sensitivity window, i.e. ${f_{\rm wide}}\equiv \sqrt{M/\pi^2 {r_{\rm wide}}^3}<f_{\rm GW}$). For $q=0.9$ and $M=60 M_\odot$, {wide nutation} of the heavier (lighter) BH can happen in the LIGO/Virgo band  if $\chi_1$ and $\chi_2$ are located in the upper-left (bottom-right) corner of the left panel of Fig.~\ref{wideconditions}. Comparable-mass BH binaries are more likely to undergo {wide nutation}, because ${r_{\rm wide}}$ increases as $q\to 1$. This is illustrated in the right panel of Fig.~\ref{wideconditions}, where we maximize the threshold ${r_{\rm wide}}$ over $\chi_1$ and $\chi_2$ -- i.e.,  we show GW frequencies corresponding to  $\max_{\chi_1<\chi_2} {r_{\rm wide}} = q^2/(1-q)^2M  $ and $\max_{\chi_2<\chi_1} {r_{\rm wide}} = 1/(1-q)^2M$. In particular, only $q\gtrsim 0.75$ allows {wide nutation} to happen in the LIGO/Virgo (LISA) band for mergers of stellar-mass (supermassive) BH binaries.  We stress the criterion  ${f_{\rm wide}}<f_{\rm GW}$ used in Fig.~\ref{wideconditions} is rather conservative, especially for low-mass systems that spend a long time in band, because the threshold ${r_{\rm wide}}$ could also be crossed later during the inspiral.

The remainder of this paper details these results. In Sec.~\ref{analytics} we derive Eq.~(\ref{rwide_intro}) analytically inside our multi-time scale framework. Sec.~\ref{numerics} presents numerical integrations to further illustrate the results and quantitatively explore the relevance of {wide nutation} in samples of BH binaries.
We conclude in Sec.~\ref{conclusions} and also point out possible directions for future work.

\section{Analytical derivation}
\label{analytics}

In this section we present analytical calculations to (i) show that {wide nutation} is indeed possible and (ii) calculate its location in the parameter space of two-spin BH binaries.

\subsection{Preliminaries}

For a quasi-circular BH binary with specified masses $m_i$,
dimensionless spins $\chi_i$ and orbital separation $r$,
the relative orientations of the Newtonian angular momentum $\mathbf{L}=m_1 m_2 \sqrt{r/M}\mathbf{\hat L}$  and the two spins $\mathbf{S_i}=m_i^2\chi_i \mathbf{\hat S_i}$  can be described using three angles (e.g. \cite{2004PhRvD..70l4020S,2010PhRvD..81h4054K,2015PhRvD..91d2003V})
\begin{align}
\cos\theta_1 = \mathbf{\hat S_1} \cdot \mathbf{\hat L} \,,
\qquad
\cos\theta_2=\mathbf{\hat S_2} \cdot \mathbf{\hat L}\,,
\qquad
\cos\Delta\Phi=\frac{\mathbf{\hat S_1} \times \mathbf{\hat L}}{|\mathbf{\hat S_1} \times  \mathbf{\hat L} |} \cdot
\frac{\mathbf{\hat S_2} \times \mathbf{\hat L}}{|\mathbf{\hat S_2} \times \mathbf{\hat L} |}\,.
\label{angdef3}
\end{align}

The timescale separation $t_{\rm pre}\ll t_{\rm rad}$ motivates the introduction of an alternative set of parameters to study BH binary dynamics~\cite{2015PhRvL.114h1103K,2015PhRvD..92f4016G},
\begin{align}
\xi = \frac{1}{M^2} \left[(1+q)\mathbf{S}_1 + \left(1+\frac{1}{q}\right)\mathbf{S}_2 \right] \cdot \hat{\mathbf{L}}\,,
\quad%
J=|\mathbf{L}+\mathbf{S_1}+\mathbf{S_2}|\,,
\quad%
S=|\mathbf{S_1}+\mathbf{S_2}|\,,
\label{xiJS}
\end{align}
which
is related to the angles of Eq.~(\ref{angdef3})
by the analytic expressions
\begin{align}
&\begin{dcases}
S = \sqrt{S_1^2 + S_2^2 + 2S_1S_2(\sin\theta_1\sin\theta_2\cos\Delta\Phi
+ \cos\theta_1\cos\theta_2)}~, \\
J = \sqrt{L^2 +  S^2 + 2L (S_1 \cos\theta_1 + S_2 \cos\theta_2)}~,\\
\xi = \frac{1+q}{q M^2}(qS_1\cos\theta_1 + S_2\cos\theta_2)~.
\end{dcases}
\label{tran1}
\end{align}

The projected effective spin $\xi$ (often denoted by $\chi_{\rm eff}$ in recent GW literature, e.g.~\cite{2016PhRvX...6d1015A})
 is conserved on both $t_{\rm pre}$ and $t_{\rm rad}$ at 2PN order \cite{2008PhRvD..78d4021R}. The magnitude of the total angular momentum $J$ remains constant on $t_{\rm pre}$ and secularly drifts on $t_{\rm rad}$. Finally, the total spin magnitude $S$ varies on the shorter time scale  $t_{\rm pre}$.  On the other hand, the three parameters $(\theta_1,\theta_2,\Delta\Phi)$ from Eq.~(\ref{angdef3}) vary on both $t_{\rm pre}$ and $t_{\rm rad}$.
Once the quantities $q$, $\chi_1$, $\chi_2$, $\xi$, $r$, and $J$ that are constant on $t_{\rm pre}$ have been fixed, the entire dynamics is parameterized by the periodic motion of $S$ \cite{2015PhRvL.114h1103K,2015PhRvD..92f4016G}.  
The magnitude $S$ oscillates from its minimum $S_-$ to its maximum $S_+$ and back to $S_-$ over its precession period $\tau$ {(which sets the timescale of the problem $t_{\rm pre} \propto r^{5/2}$)}.

Configurations where one of the two spins $\mathbf{S_i}$  is co-aligned (counter-aligned) with the orbital angular momentum $\mathbf{L}$ are described by $\theta_i\!=\!0$ ($\theta_i\!=\!\pi$). {Wide nutation} corresponds to configurations where both co-alignment and counter-alignment are allowed during the same precession cycle. The mathematical problem we wish to solve is therefore the following: \emph{can one find consistent values of the  constants of motion $(q,\chi_1,\chi_2,r, \xi,J)$ such that the binary evolution as a function of $S$ admits both $\theta_i\!=\!0$ and $\theta_i\!=\!\pi$ (for either $i=1$ or $i=2$)?}

{It is worth now differentiating between  two separate possibilities, depending on which of the two BHs ``goes wide''}. {Hereafter, we refer to the heavier (lighter) BH companion as primary (secondary).}

\subsection{{Wide nutation} of the primary black hole}
\label{sec1wide}

Let us first investigate the occurrence of {wide nutation} for the heavier BH.
By inverting Eq.~(\ref{tran1}), it is straightforward to prove that both  $\cos\theta_1$ and  $\cos\theta_2$ are monotonic in $S$ and, therefore, the functions $\theta_i(S)$ are extremized at $S=S_\pm$ (cf. Eq.~(20) in Ref.~\cite{2015PhRvD..92f4016G}). For the sake of clarity, let us rewrite Eq.~(\ref{tran1}) evaluated at these two endpoints
\begin{align}
&\begin{dcases}
S_- = \sqrt{S_1^2 + S_2^2 + 2S_1S_2(\sin\theta_{1-}\sin\theta_{2-}\cos\Delta\Phi_{-}
+ \cos\theta_{1-}\cos\theta_{2-})}~, \\
J = \sqrt{L^2 +  S_-^2 + 2L (S_1 \cos\theta_{1-} + S_2 \cos\theta_{2-})}~,\\
\xi = \frac{1+q}{q M^2}(qS_1\cos\theta_{1-} + S_2\cos\theta_{2-})~;
\end{dcases}
\label{evaluationminus}
\\
&\begin{dcases}
S_+ = \sqrt{S_1^2 + S_2^2 + 2S_1S_2(\sin\theta_{1+}\sin\theta_{2+}\cos\Delta\Phi_{+}
+ \cos\theta_{1+}\cos\theta_{2+})}~, \\
J = \sqrt{L^2 +  S_+^2 + 2L (S_1 \cos\theta_{1+} + S_2 \cos\theta_{2+})}~,\\
\xi = \frac{1+q}{q M^2}(qS_1\cos\theta_{1+} + S_2\cos\theta_{2+})~;
\end{dcases}
\label{evaluationplus}
\end{align}
where $\pm$ subscripts refer to quantities evaluated at $S_\pm$. Crucially, $J$ and $\xi$ have to be the same at $S_\pm$ because they are constant on the precession time scale, while $\theta_1$, $\theta_2$ and $\Delta\Phi$ all vary with $S$.

Because $d\cos\theta_1/dS\leq0$,
{wide nutation} is possible only if $\theta_{1-}=0$ and $\theta_{1+}=\pi$.  This yields

\begin{align}
&\begin{dcases}
J = \sqrt{ L^2 + S_1^2 + S_2^2
+ 2( L S_1
+  \cos\theta_{2-} S_1 S_2  +  \cos\theta_{2-} L S_2)}  ~,\\
\xi = \frac{1+q}{q M^2}(qS_1 + S_2\cos\theta_{2-})   ~,
\end{dcases}
\\
&\begin{dcases}
J = \sqrt{L^2 + S_1^2 + S_2^2
 - 2( L S_1
 + \cos\theta_{2+} S_1 S_2  -  \cos\theta_{2+} L S_2)} ~,\\
\xi = \frac{1+q}{q M^2}(-qS_1 + S_2\cos\theta_{2+})   ~,
\end{dcases}
\end{align}
which can be solved for $\cos\theta_{2-}$ and $\cos\theta_{2+}$. The formal solutions are
\begin{align}
\cos\theta_{2\pm}  =\frac{\pm q S_1 - L(1-q) }{S_2}\,,
\end{align}
corresponding to
\begin{align}
J = \sqrt{(1-2q) (S_1^2 - L^2) + S^2_2}\,,
\qquad
\xi = - \frac{1-q^2}{q} \frac{L}{M^2}\,.
\label{tumble1xisol}
\end{align}
These configurations are acceptable only if
$|\cos\theta_{2\pm}|\leq1$, $0<q\leq 1$, $0<\chi_i\leq 1$ and $r> M$. This is equivalent to
\begin{align}
\begin{dcases}
\chi_1\leq \chi_2 \,,
\\
r\leq {r_{\rm wide}}\,,
\end{dcases}
\label{tumblecond1}
\end{align}
where we have defined
\begin{align}
{r_{\rm wide}} \equiv \left(\frac{q \chi_2 -\chi_1}{1-q} \right)^2 M\,.
\label{rwide}
\end{align}
Any set of parameters  $(q,\chi_1,\chi_2,r)$ which satisfiy Eqs.~(\ref{tumblecond1}) therefore admits {wide nutation} for the primary BH. {Wide nutation} is realized when the remaining two constants of motions $J$ and $\xi$ assume the values given by Eq.~(\ref{tumble1xisol}).

\subsection{{Wide nutation} of the secondary black hole}
\label{sec2wide}

We now look for locations in parameter space where {wide nutation} of the secondary BH is allowed.
The calculation proceeds in a very similar fashion to the one highlighted above. The only difference is that  $d\cos\theta_2/dS\geq0$ and, therefore, one has to impose $\theta_{2-}=\pi$ and $\theta_{2+}=0$. From Eqs.~(\ref{evaluationminus}) and (\ref{evaluationplus}) we obtain

\begin{align}
&\begin{dcases}
J = \sqrt{L^2 + S_1^2 + S_2^2
- 2( L S_2
+  \cos\theta_{1-} S_1 S_2  -  \cos\theta_{1-} L S_1)}  ~,\\
\xi = \frac{1+q}{q M^2}(qS_1 \cos\theta_{1-} - S_2)   ~,
\end{dcases}
\\
&\begin{dcases}
J = \sqrt{L^2 + S_1^2 + S_2^2
 + 2( L S_2
 + \cos\theta_{1+} S_1 S_2  +  \cos\theta_{1+} L S_1)}  ~,\\
\xi = \frac{1+q}{q M^2}(qS_1 \cos\theta_{1+} + S_2)   ~,
\end{dcases}
\end{align}
which can be solved for $\cos\theta_{1-}$ and $\cos\theta_{1+}$. This yields
\begin{align}
\cos\theta_{1\pm} &=\frac{ L (1-q) \mp S_2}{q S_1}\,,
\end{align}
corresponding to
\begin{align}
J = \sqrt{\left(1-\frac{2}{q}\right) (S_2^2 - L^2) + S^2_1}\,,
\qquad
\xi =  \frac{1-q^2}{q} \frac{L}{M^2}\,.
\label{tumble2xisol}
\end{align}
The constraints $|\cos\theta_{1 \pm} |\leq 1$,
$0<q\leq 1$, $0<\chi_i\leq 1$ and $r>M$
imply
\begin{align}
&\begin{dcases}
\chi_2\leq \chi_1\,,
\\
r\leq {r_{\rm wide}}\,.
\end{dcases}\label{tumblecond2}
\end{align}

{Wide nutation} for the secondary BH is possible if Eq.~(\ref{tumblecond2}) is satisfied and it is realized for binaries described by Eq.~(\ref{tumble2xisol}).
Note that the threshold ${r_{\rm wide}}$ is the same for both Eq.~(\ref{tumblecond1}) and Eq.~(\ref{tumblecond2}), %
and the first condition involving the spin magnitudes is symmetric under the exchange $1\!\xleftrightarrow\!2$.

\subsection{Remarks}
\label{remarks}

The derivation we have just presented highlights the occurrence of the separation threshold ${r_{\rm wide}}$. Below ${r_{\rm wide}}$, PN spin-spin and spin-orbit couplings are strong enough to allow for {wide nutation}.
Ref.~\cite{2015PhRvL.115n1102G} located two further separation thresholds in the  parameter space of double-spinning BH binaries, namely
\begin{align}
r_{\rm ud\pm} = \frac{(\sqrt{\chi_1} \pm \sqrt{q \chi_2})^4}{(1-q)^2}~M\,.
\end{align}
In the range $r_{\rm ud-}<r<r_{\rm ud+}$,
a precessional instability affects binaries where the primary BH spin is co-aligned and the secondary BH spin is counter-aligned with the orbital angular momentum. It is straightforward to prove that ${r_{\rm wide}}=\sqrt{r_{\rm ud-}\, r_{\rm ud+}}$ and therefore $r_{\rm ud-} \leq {r_{\rm wide}} \leq r_{\rm ud+}$. Given an ensemble of binaries with the same $(q,\,\chi_1,\,\chi_2)$ inspiraling from large to small separations, the up-down configuration becomes unstable {before} any other binary of the ensemble is allowed to undergo {wide nutation}.

In the equal-mass limit $q\rightarrow 1$, ${r_{\rm wide}}$ evidently
diverges,
such that the radial conditions of Eq.~(\ref{tumblecond1}) and (\ref{tumblecond2}) are always satisfied (one can prove this remains the case even when $\chi_1=\chi_2$).
For $q=1$, $S=[J^2 - L^2 - L \xi M^2]^{1/2}$ becomes a constant of motion on both $t_{\rm pre}$ and $t_{\rm rad}$ \cite{2017CQGra..34f4004G}. Evaluating Eq.~(\ref{tumble1xisol}) and Eq.~(\ref{tumble2xisol}) for $q=1$ yields $J = [\, L^2 + | S_1^2 - S_2^2 | \,]^{1/2}$,  $\xi = 0$
and thus $S=|S_1^2 - S_2^2|$. This in agreement with Eqs.~(2.25) and (2.26) of Ref.~\cite{2017CQGra..34f4004G}. For $q=1$, if the condition for {wide nutation} is met, it holds for
{\em all} values of $r$ throughout the entire inspiral. This observation
also demonstrates that {wide nutation} is not necessarily a phenomenon
encountered only at small binary separations.

\section{Numerical explorations}
\label{numerics}

We now test our findings against numerical integrations. All PN integrations presented in this paper are performed using the \textsc{precession} code \cite{2016PhRvD..93l4066G}. Spin-precession equations are accurate up to 2PN. Orbit-averaged radiation reaction is computed including
(non)spinning terms up to (3.5PN) 2PN. %
 Precession-averaged evolutions of $J(r)$ on $t_{\rm rad}$ \cite{2015PhRvL.114h1103K,2015PhRvD..92f4016G} are accurate up to 1PN.

\subsection{{Wide nutation} phenomenology}

\begin{figure*}[t!]
\centering
\includegraphics[width=0.49\textwidth,page=1]{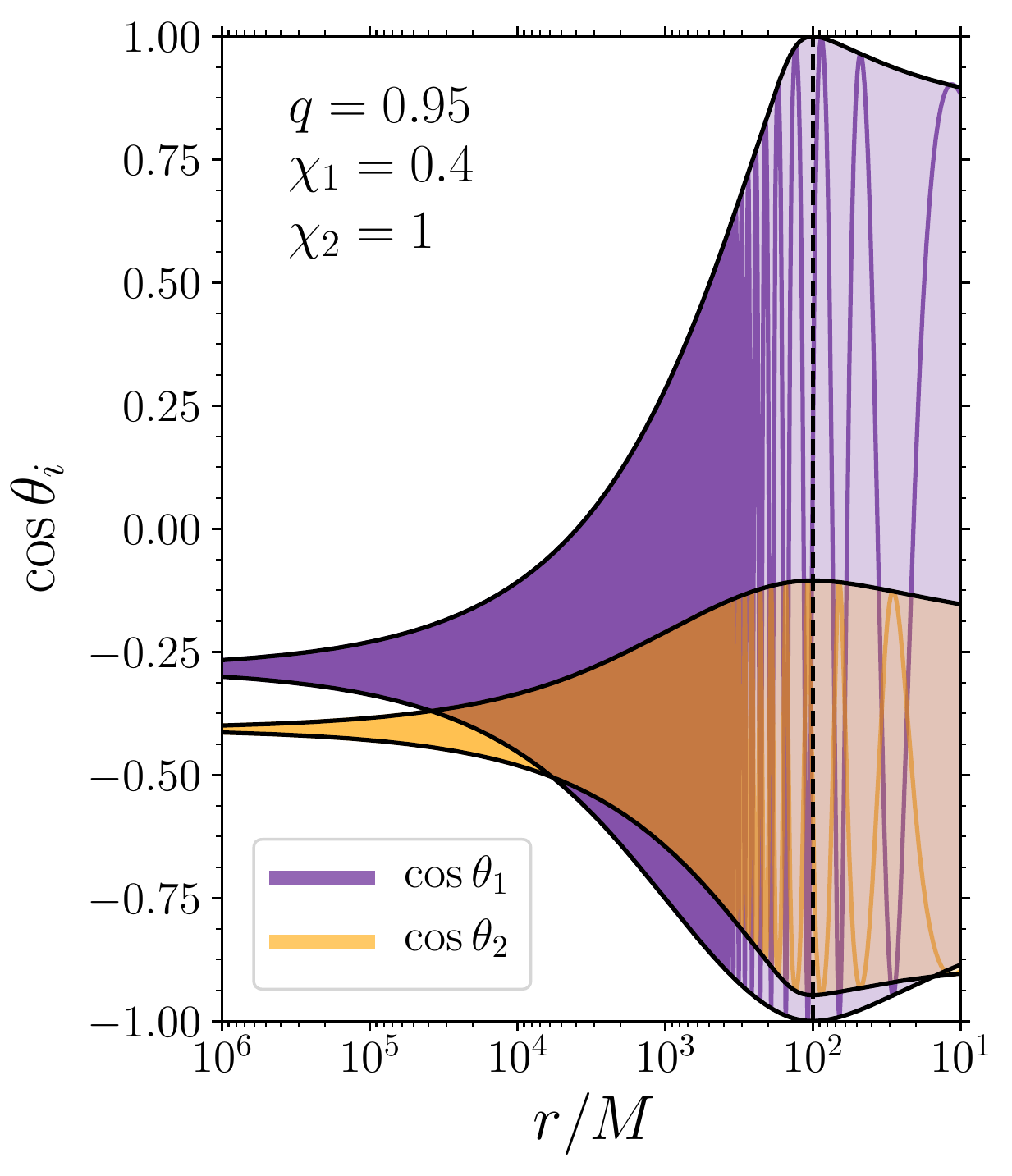}
\includegraphics[width=0.49\textwidth,page=2]{envelopes}
\caption{PN evolutions of the tilt angles for {wide binaries}. Left (right) panels show cases where the primary (secondary) BH spin undergoes {wide  nutation}. Binaries are characterized by $q=0.95,\chi_1=1, \chi_2=0.4$ (left) and  $q=0.95,\chi_1=0.4, \chi_2=1$ (left); the values of $J$ and $\xi$ are chosen such that {wide  nutation} happens at $r=100M<{r_{\rm wide}}$ (dashed line). Shaded regions show precession-averaged evolutions of the tilt angles $\theta_1$ (purple) and $\theta_2$ (orange). Their envelopes (black lines) correspond to $\cos\theta_{i \pm}$, as defined in Sec.~\ref{analytics}. The oscillatory lines show a representative orbit-averaged evolution obtained after sampling the precessional phase at $r=10^6 M$.}
\label{envelopes}
\end{figure*}

The orbit-averaged PN inspiral of two representative binaries undergoing {wide nutation} is shown in Fig.~\ref{envelopes}. The parameters of the binary illustrated in the left (right) panel are chosen such that the primary (secondary) BH undergoes {wide nutation} at $r=100 M<{r_{\rm wide}}$. Because both $J$ and $r$ evolve on $t_{\rm rad}$,  binaries stay in the {wide nutation} configuration only at the specific separation where either one of Eq.~(\ref{tumble1xisol}) or Eq.~(\ref{tumble2xisol}) are satisfied.
As the {wide-nutation} separation is approached, oscillations of either of the two spins grow larger and larger. The envelopes reach their extreme values $|\cos\theta_{i+} - \cos\theta_{i-}| = 2$ at $r=100M$. The time scale hierarchy $t_{\rm pre}\ll t_{\rm rad}$ ensures that
 {wide nutation} corresponds to multiple cycles with large spin oscillations.   For separations $r\sim100 M$, the {wide spin} repeatedly flips over a large range of width $\delta\theta_i\lesssim \pi$.

\begin{figure*}[t!]
\centering
\includegraphics[width=0.49\textwidth,page=1]{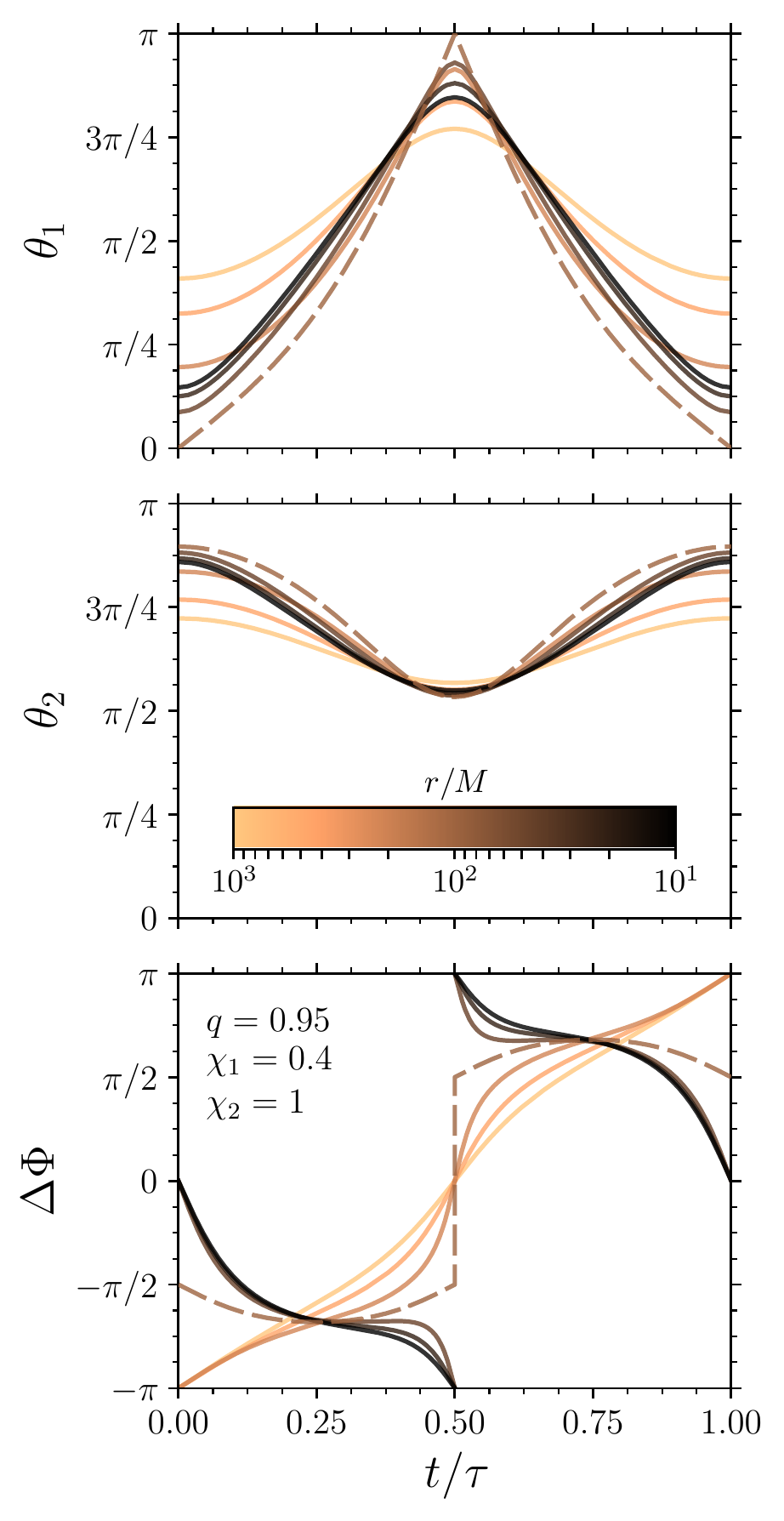}
\includegraphics[width=0.49\textwidth,page=2]{cyclesevol}
\caption{Evolution of $\theta_1$ (top), $\theta_2$ (middle) and $\Delta\Phi$ (bottom) for the same {wide binaries} of Fig.~\ref{envelopes}. Left (right) panels show a BH binary where the primary (secondary) spin undergoes {wide  nutation} at $r=100M$. Light (dark) curves correspond to large (small) separations. {Wide  nutation} is marked with a dashed line. Binaries evolve along those curves on $t_{\rm pre}$, while the curves themselves evolve on $t_{\rm rad}$. An animated version of this figure is available at \href{https://davidegerosa.com/spinprecession/}{www.davidegerosa.com/spinprecession}.}
\label{cyclesevol}
\end{figure*}

Fig.~\ref{cyclesevol} shows precession-averaged time evolutions of the angles $\theta_1$, $\theta_2$ and $\Delta\Phi$ for the same representative binaries. Each of the curves describes a full precession cycle at fixed orbital separation: binaries start at $t=0$ with $S=S_-$, evolve to $S=S_+$ at $t=\tau/2$, and return to $S=S_-$ after a period $\tau$. Binaries evolve along these curves on the short time scale $t_{\rm pre}$, while the curves themselves evolve secularly on $t_{\rm rad}$.

The spin tilts $\theta_i$ approach the {wide nutation} configuration rather gently, without abrupt changes in their evolution.
In contrast, the evolution $\Delta\Phi(t)$ is discontinuous at the {wide-nutation} separation. {Wide nutation} is characterized by $\sin\theta_{i\pm}\!=\!0$ and, consequently, discontinuities in $\Delta\Phi$ are expected at $S_\pm$ (i.e. $t=0,\tau/2,\tau$)  [cf. Eq.~(\ref{angdef3})]. Figure~\ref{cyclesevol} indicates that $\lim _{S\to S_\pm} |\Delta\Phi|= \pi/2$, which can be understood as follows. Imagine approaching the Earth's pole along a meridian: the longitude is not defined at the pole and discontinuously changes by $\pi$ as the pole is crossed. Similarly, $\Delta\Phi$ is not defined at $S_\pm$ for {wide binaries} and the discontinuity jump across those points is equal to $\pi$. Because the evolution of $\Delta\Phi$ is odd about $t=\tau/2$ \cite{2015PhRvD..92f4016G}, one thus needs $\lim _{S\to S_\pm} |\Delta\Phi|= \pi/2$.%

For {widely nutating} binaries, one of the two spins touches both a completely aligned and a completely anti-aligned configuration during a single precession period $\tau$. Oscillations of the polar angles $\theta_i$ are nutations: {these solutions} correspond to maximal nutations in a non-inertial frame which co-precesses with $\mathbf{L}$. In an inertial frame, on the other hand, one needs to consider the motion of $\mathbf{L}$ about $\mathbf{J}$ (whose direction is approximately constant \cite{2017PhRvD..96b4007Z}). This can be described by a polar angle $\theta_L$ and an azimuthal angle $\alpha$ \cite{2015PhRvD..92f4016G}. For the cases shown in Figs.~\ref{envelopes} and \ref{cyclesevol}, $\mathbf{L}$ precesses about $\mathbf{J}$ about $\alpha/2\pi\!\sim\!18$ times during each period $\tau$ at  $r \!\sim\!100M$. At the same time, $\theta_L$ changes by $\sim 3^\circ$.

\subsection{Almost {wide binaries}}

\begin{figure*}[t!]
\centering
\includegraphics[width=\textwidth,page=1]{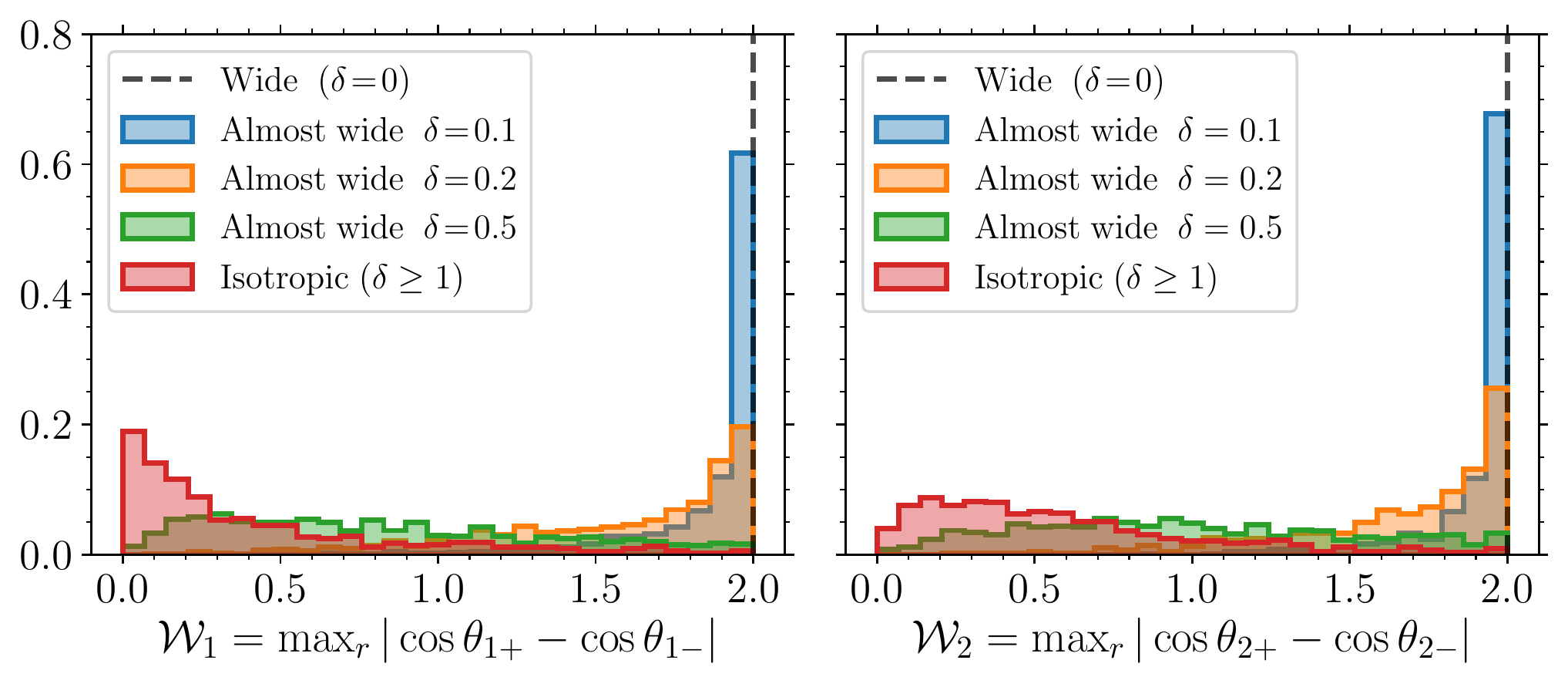}
\caption{Histograms of the estimator $\mathcal{W}_i=\max_{r} |\cos\theta_{i+}-\cos\theta_{i-}|$ for {wide, almost wide} and isotropic binary BHs. Left (right) panel considers {wide nutation} of the more (less) massive BH. The parameter $\delta$ measures the proximity of the configurations considered to {wide nutation}. {Wide} binaries have $\delta=0$ and $\mathcal{W}_i=2$.}
\label{almostwide}
\end{figure*}

The {wide-nutation} conditions of Sec.~\ref{analytics} are  fine tuned. While generic binaries satisfy the inequalities~(\ref{tumblecond1}) and (\ref{tumblecond2}), only an infinitesimally small set of  configurations satisfies Eqs.~(\ref{tumble1xisol}) and (\ref{tumble2xisol}) for $J$ and $\xi$. It is natural, therefore, to address the behavior of \emph{{almost wide}} binaries, i.e. sources which satisfy the {wide-nutation} conditions only approximately. This point is explored in Fig.~\ref{almostwide}, where we evaluate the estimator
\begin{equation}
\mathcal{W}_i \big(q,\chi_1,\chi_2,\theta_{1\infty},\theta_{2\infty}\big)=\max_{r} |\cos\theta_{i+}-\cos\theta_{i-}|\,.
\end{equation}
One has $0\leq \mathcal{W}_i \leq 2$ and $\mathcal{W}_i=2$ corresponds to {wide nutation} (cf. the envelopes in Fig.~\ref{envelopes}).  Because the angles $\theta_i$ are asymptotically constant (for $q<1$), generic binaries satisfy $\lim_{r/M\to\infty}  |\cos\theta_{i+}-\cos\theta_{i-}| =0$.

For a statistical analysis, we generate a sample of binaries uniformly in $q,\chi_1,\chi_2\in [0.1,1]$ and  $\log (r/M) \in [6,1]$. We first filter the sample according to the inequalities (\ref{tumblecond1}) ($i=1$, left panel) and (\ref{tumblecond2}) ($i=2$, right panel). This leaves about $1\%$ of the sources. We consider (exactly) {wide binaries} from Eqs.~(\ref{tumble1xisol}) and (\ref{tumble2xisol}) and evolve them backwards in time to $r/M\!\to\!\infty$ obtaining values of $\theta_{1\infty}$ and $\theta_{2\infty}$~\cite{2015PhRvD..92f4016G}. If these asymptotic configurations are evolved back down to $r\!=\!10M$, we trivially obtain $\mathcal{W}_i=2$ (gray dashed lines in Fig.~\ref{almostwide}). The same systems are then perturbed by substituting $q$, $\chi_i$ and $\cos\theta_{i\infty}$ to new values generated uniformly in $[q-\delta, q+\delta]$, $[\chi_i-\delta, \chi_i+\delta]$, and $[\cos\theta_{i\infty}-2\delta, \cos\theta_{i\infty}+2\delta]$, where $\delta$ is a constant (random draws are restricted to the intervals $q,\chi_i \in [0.1,1]$ and $\cos\theta_i\in[-1,1]$). In this series of distributions, the parameter $\delta$ measures the ``distance from {wideness}'':  $\delta=0$ corresponds to exactly {wide} binaries, while $\delta\geq 1$ corresponds to the (unfiltered) starting sample of binaries with isotropic spin directions. 
Fig.~\ref{almostwide} shows that these \emph{almost {wide}} sources present $\mathcal{W}_i\gtrsim 1.5$ even for moderately large values $\delta\lesssim 0.2$. For comparison, the isotropic distribution presents $\mathcal{W}_i\lesssim 0.5$.

Despite being a fine-tuned configuration, {wide nutation} appears to be the extreme limit of a more generic class of binaries with similar phenomenology. In other terms, \emph{almost {wide}} binaries are still very {wide}.

\subsection{Relation to the spin morphologies}

\begin{figure*}[t!]
\centering
\includegraphics[width=0.495\textwidth,page=1]{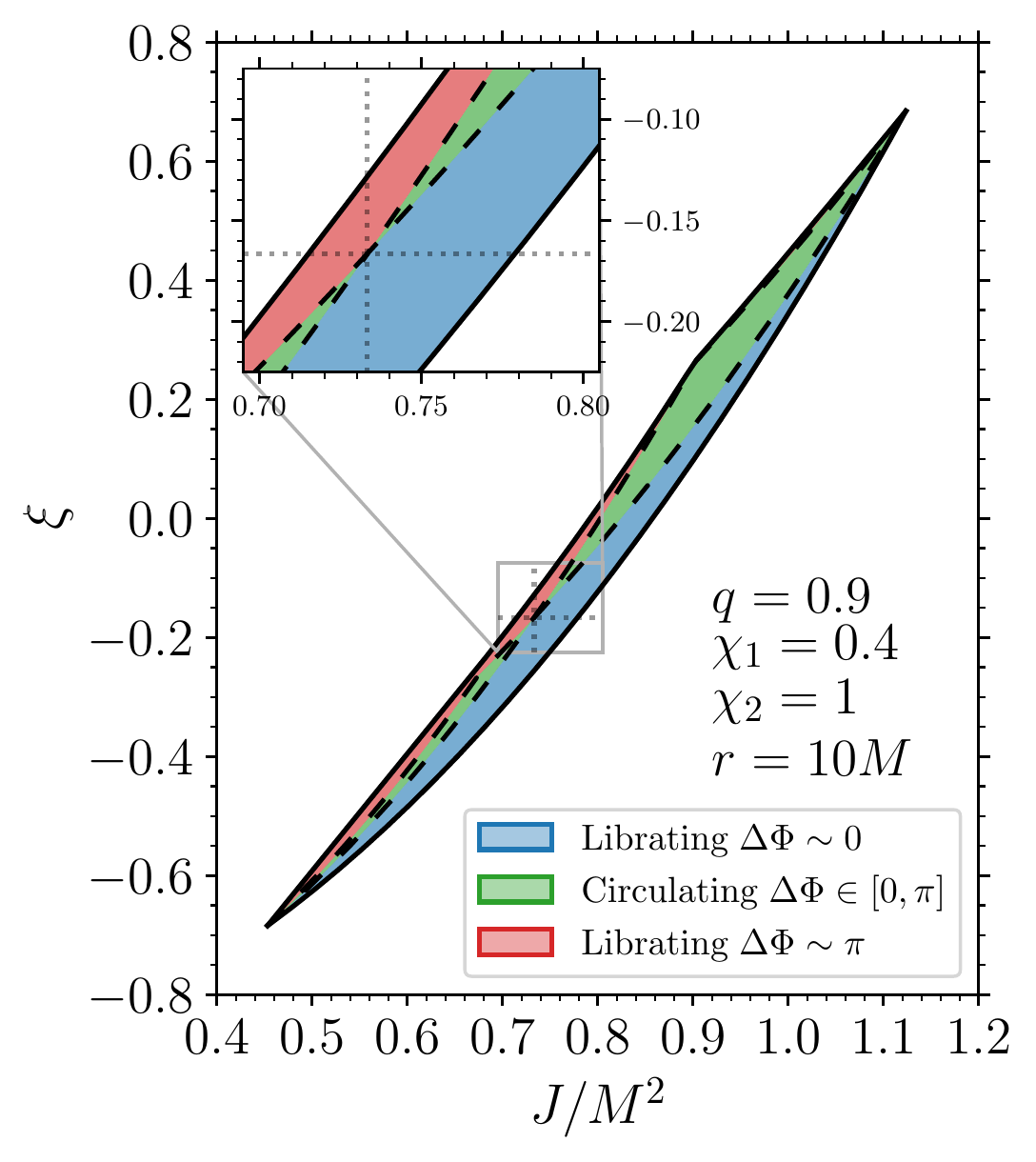}
\includegraphics[width=0.495\textwidth,page=2]{Jxiplane}
\caption{Allowed region in the $J\!-\!\xi$ plane for sets of parameters allowing for {wide  nutation} of the primary (left) and secondary (right) BHs. Solid and dashed black line mark the location of the spin-orbit resonances and morphological transitions, respectively. The colored regions correspond to the different $\Delta\Phi$ morphologies. {Wide  nutation} is located at the intersection of the two gray dotted lines, where  three colored regions and two dashed lines all meet.  }
\label{Jxiplane}
\end{figure*}

In Refs.~\cite{2015PhRvL.114h1103K,2015PhRvD..92f4016G}, the dynamics of precessing BH binaries was classified in terms of  the evolution of $\Delta\Phi$ on $t_{\rm pre}$. Three discrete cases (or \emph{morphologies}) are possible, where $|\Delta\Phi|$ either circulates in the full range $[0,\pi]$ (i.e. either $|\Delta\Phi_+|\!=\!0$ and $|\Delta\Phi_-|\!=\!\pi$, or $|\Delta\Phi_+|\!=\!\pi$ and $|\Delta\Phi_-|\!=\!0$), librates about $|\Delta\Phi_\pm|\!=\!0$ and never reaches $\pi$, or librates about $|\Delta\Phi_\pm|\!=\!\pi$ and never reaches $0$. One has $\lim_{r/M\to\infty}  |\Delta\Phi_-|\!=\!\pi$ and $\lim_{r/M\to\infty}   |\Delta\Phi_+|\!=\!0$ \cite{2015PhRvD..92f4016G}, so all binaries belong to the circulating morphology at large separations where they form (cf. also Fig.~\ref{cyclesevol} above). Secular evolution under radiation reaction generically induces transitions between the different morphologies.

Fig.~\ref{Jxiplane} shows the $(J\!-\!\xi)$ parameter space of BH binaries for values of $q,\chi_1,\chi_2$ and $r$ which admit {wide nutation}. Geometrical constraints imply that binaries can only ``live'' within the wedge bounded by the solid black lines. The allowed area is colored according to the $\Delta\Phi$ spin morphologies. The two librating morphologies are located close to the edges of the parameter space (the edges themselves are  the so-called spin-orbit resonances of Ref.~\cite{2004PhRvD..70l4020S}), while circulating binaries are found in between.  Radiation reaction causes binaries to move along horizontal lines in Fig.~\ref{Jxiplane} (because $\xi$ is constant). In particular, they can cross the dashed lines and move into different colored regions.%

{Wide-nutation} binaries are located at the special points of these diagrams where all three colored regions meet. Such locations do not exist in general and are only allowed if the conditions derived in Sec.~\ref{analytics} are met. As shown in Fig.~\ref{cyclesevol} above, binaries evolving across {wide nutation} remain in the circulating morphology and transition from ($|\Delta\Phi_-|\!=\!\pi$, $|\Delta\Phi_-|\!=\!0$) to ($|\Delta\Phi_-|\!=\!0$, $|\Delta\Phi_-|\!=\!\pi$).  Morphological transitions occur when radiation reaction changes $r$ and $J$ to values at which $\theta_i = 0$ or $\pi$ at $S_-$ or $S_+$ \cite{2015PhRvD..92f4016G}.  
A {wide} binary can therefore be interpreted as the occurrence of two morphological transitions at the same value of $r$, with radiation reaction simultaneously driving $\theta_i$ to $0$ at one extremum of $S$ and to $\pi$ at the other.

These findings seem to show that {wide nutation} somehow resembles triple points in phase-space diagrams of thermodynamical systems. The reformulation of BH spin-precession dynamics using thermodynamical arguments is a promising avenue for future work.

\section{Conclusions}
\label{conclusions}

In generic spinning BH binaries, both the orbital plane and the spins precess around the total angular momentum. The simple scenario (often called \emph{simple precession} \cite{1994PhRvD..49.6274A}), where the two spins evolve over cones of fixed amplitude $\theta_1$ and $\theta_2$ about the total angular momentum, is strictly valid only at large separations. Closer to merger spin-spin couplings become important, and BH spins undergo complicated precessional and nutational motions. The tilt angles $\theta_i$, as well as the angle $\Delta\Phi$ between the projections of the spins on the orbital plane, change on the short precession timescale. Here we pointed out the existence of configurations where nutations are maximized: in these BH binaries, one of the two spins oscillates from $\theta_i=0$ to $\theta_i=\pi$ {\em during a single precession cycle.}

For {wide nutation} to be possible, the spins must satisfy the conditions $\chi_1\leq \chi_2$ (for the primary), $\chi_2\leq \chi_1$ (for the secondary), and the binary separation $r$ must be smaller than ${r_{\rm wide}}\equiv M (q \chi_2 -\chi_1)^2/ (1-q)^2$. Therefore, {wide nutation} is easier to achieve if the binary's mass ratio is close to unity ($q\lesssim 1$) and/or if the two spin magnitudes are very different from each other ($\chi_1\ll \chi_2$ or $\chi_2\ll \chi_1$). 
If allowed, {wide nutation} happens at the locations in parameter space given by Eqs.~(\ref{tumble1xisol}) and (\ref{tumble2xisol}). In particular, the effective spin $\xi\equiv \chi_{\rm eff}$ must be close to zero. Interestingly, the conditions $q\lesssim 1$ and $\xi\sim 0$ are met by most of the LIGO/Virgo events observed to date \cite{2016PhRvX...6d1015A}. {This behavior was derived exploiting the precession-averaged formalism \cite{2015PhRvL.114h1103K,2015PhRvD..92f4016G}. Our approach is expected to break down only at separations $r\sim 10M$ \cite{2015PhRvD..92f4016G}, where the PN approximation becomes
inaccurate for binaries of comparable masses \cite{2006PhRvD..74j4005B,2009PhRvD..79h4010C,2009PhRvD..80h4043B}.}

Moreover, {wide nutation} turns out to be just the tip of the iceberg of a much wider class of \emph{almost {wide}} binaries. %
 About $\sim 30\%$ of binaries with $q > 0.8$ and $|\xi| < 0.1$ will experience oscillations  in either $\cos \theta_1$  or $\cos\theta_2$ larger than $1.5$ at some point in their inspiral\footnote{This fraction is estimated assuming uniform distributions in $\chi_i\in[0.1,1]$ and isotropic spin orientations.}. For these binaries, nutations are maximized in the late inspiral ($r \sim 10 M$), which is potentially observable by current and future detectors.

Hints of these configurations have been previously found in Refs.~\cite{2015PhRvL.114n1101L,2016PhRvD..93d4031L}, but here we have generalized those findings and made them more rigorous by separating the timescales of the problem. In particular, Ref.~\cite{2015PhRvL.114n1101L} presented a BH-binary numerical-relativity simulation where one of the two spins has a large nutation (or ``{flip flop}'' in the language of Refs.~\cite{2015PhRvL.114n1101L,2016PhRvD..93d4031L}). The simulation parameters were chosen such that $q=1$, $\mathbf{S_1}\times \mathbf{L} =0$ and $\mathbf{S} \cdot \mathbf{L}= 0$. These choices correspond to $\xi=0$ and $S=|S_1^2-S_2^2|$, and turn out to be equivalent to our solutions when restricted to equal-mass binaries (cf. Sec.~\ref{remarks}). ``{Flip flops}'' were later investigated in Ref.~\cite{2016PhRvD..93d4031L} using PN evolutions. They  found the occurrence of a critical separation below which large nutations are possible; their expression, however, is different and more complicated than our Eq.~(\ref{rwide}). The ``{flip flop}'' angle of the primary BH is maximized by imposing alignment of $\mathbf{S_1}$ and $\mathbf{L}$, while anti-alignment is required in the case of the secondary BH \cite{2016PhRvD..93d4031L}. However, {both} alignment {and} anti-alignment must be imposed at the same time for nutations to be maximal, by definition (cf. Sec.~\ref{sec1wide}-\ref{sec2wide}).  The conditions $\chi_1 \gtrless \chi_2$ of Eqs.~(\ref{tumblecond1}) and (\ref{tumblecond2}) do not appear to be mentioned, at least explicitly, in Ref.~\cite{2016PhRvD..93d4031L}.

  Future work on this subject could target {widely nutating} binaries with numerical-relativity simulations (cf. \cite{2015PhRvD..92j4028O}) and assess their measurability through injections in LIGO/Virgo parameter-estimation pipelines. {Measuring spin dynamics with current detectors remains challenging \cite{2016PhRvD..93h4042P,2018arXiv180910113K}; work is in progress to accurately forecast the measurability of two-spin effects, like wide nutation, with future instruments.} The upcoming age of high-precision GW astronomy will soon uncover many new phenomena. As shown in this paper, BH binary spin {dynamics} might still have some surprises in store.

\section*{Acknowledgements }
We thank Yanbei Chen, Alan Weinstein and Kaze Wong for discussions.
D.G. is supported by NASA through Einstein Postdoctoral Fellowship Grant No. PF6-170152 by the Chandra X-ray Center, operated by the Smithsonian Astrophysical Observatory for NASA under Contract NAS8-03060.
A.L. is supported by the LIGO SURF program at Caltech funded by NSF Grant No. PHY-1757303 and the NSBP Carl Albert Rouse Undergraduate Research Fellowship.
E.B. is supported by NSF Grants Nos. PHY-1841464, AST-1841358 and PHY-090003, and NASA Grant No. 17-ATP17-0225. M.K. is supported by NSF Grant No. PHY-1607031.
U.S. acknowledges support by
European Union's H2020 
ERC 
Grant
No. MaGRaTh--646597, Marie Sk\l{}odowska-Curie Grant  No. 690904,
COST Action Grant No.~CA16104, 
the Yukawa Institute for Theoretical Physics at Kyoto
University under Grant No. YITP-T-18-05,
STFC 
Grant Nos.~ST/P000673/1, ST/H008586/1 and ST/K00333X/1,
 PRACE Grant No.~2016163948, BIS Grant No.~ST/J005673/1 and NSF Grant No.~PHY-090003.
Computational
work was performed on Caltech computer cluster
\emph{Wheeler} supported by the Sherman Fairchild Foundation and Caltech.

\section*{References}
\bibliographystyle{iopart-num_leo}
\bibliography{wideprecession}

\providecommand{\newblock}{}
\begin{thebibliography}{10}
\expandafter\ifx\csname url\endcsname\relax
  \def\url#1{{\tt #1}}\fi
\expandafter\ifx\csname urlprefix\endcsname\relax\def\urlprefix{URL }\fi
\providecommand{\eprint}[2][]{\href{http://arxiv.org/abs/#2}{arXiv:#2}}

\bibitem{2014PhRvL.112y1101V}
{Vitale} S, {Lynch} R, {Veitch} J, {Raymond} V and {Sturani} R 2014
  \href{http://dx.doi.org/10.1103/PhysRevLett.112.251101}{ {\em \prl\/} {\bf
  112} 251101 } [\eprint{1403.0129}]

\bibitem{2016PhRvD..93d4071T}
{Trifir{\`o}} D, {O'Shaughnessy} R, {Gerosa} D, {Berti} E, {Kesden} M,
  {Littenberg} T and {Sperhake} U 2016
  \href{http://dx.doi.org/10.1103/PhysRevD.93.044071}{ {\em \prd\/} {\bf 93}
  044071 } [\eprint{1507.05587}]

\bibitem{2017PhRvD..96l4041W}
{Williamson} A~R, {Lange} J, {O'Shaughnessy} R, {Clark} J~A, {Kumar} P,
  {Calder{\'o}n Bustillo} J and {Veitch} J 2017
  \href{http://dx.doi.org/10.1103/PhysRevD.96.124041}{ {\em \prd\/} {\bf 96}
  124041 } [\eprint{1709.03095}]

\bibitem{1994PhRvD..49.2658C}
{Cutler} C and {Flanagan} {\'E}~E 1994
  \href{http://dx.doi.org/10.1103/PhysRevD.49.2658}{ {\em \prd\/} {\bf 49}
  2658--2697 } [\eprint{gr-qc/9402014}]

\bibitem{2015ApJ...798L..17C}
{Chatziioannou} K, {Cornish} N, {Klein} A and {Yunes} N 2015
  \href{http://dx.doi.org/10.1088/2041-8205/798/1/L17}{ {\em \apjl\/} {\bf 798}
  L17 } [\eprint{1402.3581}]

\bibitem{2016ApJ...832L...2R}
{Rodriguez} C~L, {Zevin} M, {Pankow} C, {Kalogera} V and {Rasio} F~A 2016
  \href{http://dx.doi.org/10.3847/2041-8205/832/1/L2}{ {\em \apjl\/} {\bf 832}
  L2 } [\eprint{1609.05916}]

\bibitem{2017Natur.548..426F}
{Farr} W~M, {Stevenson} S, {Miller} M~C, {Mandel} I, {Farr} B and {Vecchio} A
  2017 \href{http://dx.doi.org/10.1038/nature23453}{ {\em \nat\/} {\bf 548}
  426--429 } [\eprint{1706.01385}]

\bibitem{2017PhRvD..95l4046G}
{Gerosa} D and {Berti} E 2017
  \href{http://dx.doi.org/10.1103/PhysRevD.95.124046}{ {\em \prd\/} {\bf 95}
  124046 } [\eprint{1703.06223}]

\bibitem{2018ApJ...862L...3S}
{Schr{\o}der} S~L, {Batta} A and {Ramirez-Ruiz} E 2018
  \href{http://dx.doi.org/10.3847/2041-8213/aacf8d}{ {\em \apjl\/} {\bf 862} L3
  } [\eprint{1805.01269}]

\bibitem{2017PhRvL.119a1101O}
{O'Shaughnessy} R, {Gerosa} D and {Wysocki} D 2017
  \href{http://dx.doi.org/10.1103/PhysRevLett.119.011101}{ {\em \prl\/} {\bf
  119} 011101 } [\eprint{1704.03879}]

\bibitem{2018PhRvD..97d3014W}
{Wysocki} D, {Gerosa} D, {O'Shaughnessy} R, {Belczynski} K, {Gladysz} W,
  {Berti} E, {Kesden} M and {Holz} D~E 2018
  \href{http://dx.doi.org/10.1103/PhysRevD.97.043014}{ {\em \prd\/} {\bf 97}
  043014 } [\eprint{1709.01943}]

\bibitem{2017arXiv170607053B}
{Belczynski} K, {Klencki} J, {Meynet} G, {Fryer} C~L, {Brown} D~A,
  {Chruslinska} M, {Gladysz} W, {O'Shaughnessy} R, {Bulik} T, {Berti} E, {Holz}
  D~E, {Gerosa} D, {Giersz} M, {Ekstrom} S, {Georgy} C, {Askar} A, {Wysocki} D
  and {Lasota} J~P 2017  [\eprint{1706.07053}]

\bibitem{2019MNRAS.483.3288P}
{Postnov} K~A and {Kuranov} A~G 2019
  \href{http://dx.doi.org/10.1093/mnras/sty3313}{ {\em \mnras\/} {\bf 483}
  3288--3306 } [\eprint{1706.00369}]

\bibitem{2013PhRvD..87j4028G}
{Gerosa} D, {Kesden} M, {Berti} E, {O'Shaughnessy} R and {Sperhake} U 2013
  \href{http://dx.doi.org/10.1103/PhysRevD.87.104028}{ {\em \prd\/} {\bf 87}
  104028 } [\eprint{1302.4442}]

\bibitem{2018PhRvD..98h4036G}
{Gerosa} D, {Berti} E, {O'Shaughnessy} R, {Belczynski} K, {Kesden} M, {Wysocki}
  D and {Gladysz} W 2018 \href{http://dx.doi.org/10.1103/PhysRevD.98.084036}{
  {\em \prd\/} {\bf 98} 084036 } [\eprint{1808.02491}]

\bibitem{1994PhRvD..49.6274A}
{Apostolatos} T~A, {Cutler} C, {Sussman} G~J and {Thorne} K~S 1994
  \href{http://dx.doi.org/10.1103/PhysRevD.49.6274}{ {\em \prd\/} {\bf 49}
  6274--6297 }

\bibitem{2004PhRvD..70l4020S}
{Schnittman} J~D 2004 \href{http://dx.doi.org/10.1103/PhysRevD.70.124020}{ {\em
  \prd\/} {\bf 70} 124020 } [\eprint{astro-ph/0409174}]

\bibitem{2007PhRvL..98w1102C}
{Campanelli} M, {Lousto} C~O, {Zlochower} Y and {Merritt} D 2007
  \href{http://dx.doi.org/10.1103/PhysRevLett.98.231102}{ {\em \prl\/} {\bf 98}
  231102 } [\eprint{gr-qc/0702133}]

\bibitem{2008PhRvD..77l4047B}
{Br{\"u}gmann} B, {Gonz{\'a}lez} J~A, {Hannam} M, {Husa} S and {Sperhake} U
  2008 \href{http://dx.doi.org/10.1103/PhysRevD.77.124047}{ {\em \prd\/} {\bf
  77} 124047 } [\eprint{0707.0135}]

\bibitem{2010ApJ...715.1006K}
{Kesden} M, {Sperhake} U and {Berti} E 2010
  \href{http://dx.doi.org/10.1088/0004-637X/715/2/1006}{ {\em \apj\/} {\bf 715}
  1006--1011 } [\eprint{1003.4993}]

\bibitem{2011PhRvL.107w1102L}
{Lousto} C~O and {Zlochower} Y 2011
  \href{http://dx.doi.org/10.1103/PhysRevLett.107.231102}{ {\em \prl\/} {\bf
  107} 231102 } [\eprint{1108.2009}]

\bibitem{2018PhRvD..97j4049G}
{Gerosa} D, {H{\'e}bert} F and {Stein} L~C 2018
  \href{http://dx.doi.org/10.1103/PhysRevD.97.104049}{ {\em \prd\/} {\bf 97}
  104049 } [\eprint{1802.04276}]

\bibitem{2015PhRvL.114n1101L}
{Lousto} C~O and {Healy} J 2015
  \href{http://dx.doi.org/10.1103/PhysRevLett.114.141101}{ {\em \prl\/} {\bf
  114} 141101 } [\eprint{1410.3830}]

\bibitem{2016PhRvD..93d4031L}
{Lousto} C~O, {Healy} J and {Nakano} H 2016
  \href{http://dx.doi.org/10.1103/PhysRevD.93.044031}{ {\em \prd\/} {\bf 93}
  044031 } [\eprint{1506.04768}]

\bibitem{1964PhRv..136.1224P}
{Peters} P~C 1964 \href{http://dx.doi.org/10.1103/PhysRev.136.B1224}{ {\em
  Physical Review\/} {\bf 136} 1224--1232 }

\bibitem{2008PhRvD..78d4021R}
{Racine} {\'E} 2008 \href{http://dx.doi.org/10.1103/PhysRevD.78.044021}{ {\em
  \prd\/} {\bf 78} 044021 } [\eprint{0803.1820}]

\bibitem{2015PhRvL.114h1103K}
{Kesden} M, {Gerosa} D, {O'Shaughnessy} R, {Berti} E and {Sperhake} U 2015
  \href{http://dx.doi.org/10.1103/PhysRevLett.114.081103}{ {\em \prl\/} {\bf
  114} 081103 } [\eprint{1411.0674}]

\bibitem{2015PhRvD..92f4016G}
{Gerosa} D, {Kesden} M, {Sperhake} U, {Berti} E and {O'Shaughnessy} R 2015
  \href{http://dx.doi.org/10.1103/PhysRevD.92.064016}{ {\em \prd\/} {\bf 92}
  064016 } [\eprint{1506.03492}]

\bibitem{2017PhRvD..95j4004C}
{Chatziioannou} K, {Klein} A, {Yunes} N and {Cornish} N 2017
  \href{http://dx.doi.org/10.1103/PhysRevD.95.104004}{ {\em \prd\/} {\bf 95}
  104004 } [\eprint{1703.03967}]

\bibitem{2017PhRvL.118e1101C}
{Chatziioannou} K, {Klein} A, {Cornish} N and {Yunes} N 2017
  \href{http://dx.doi.org/10.1103/PhysRevLett.118.051101}{ {\em \prl\/} {\bf
  118} 051101 } [\eprint{1606.03117}]

\bibitem{2018arXiv180910113K}
{Khan} S, {Chatziioannou} K, {Hannam} M and {Ohme} F 2018
  [\eprint{1809.10113}]

\bibitem{2017PhRvD..96b4007Z}
{Zhao} X, {Kesden} M and {Gerosa} D 2017
  \href{http://dx.doi.org/10.1103/PhysRevD.96.024007}{ {\em \prd\/} {\bf 96}
  024007 } [\eprint{1705.02369}]

\bibitem{2015PhRvL.115n1102G}
{Gerosa} D, {Kesden} M, {O'Shaughnessy} R, {Klein} A, {Berti} E, {Sperhake} U
  and {Trifir{\`o}} D 2015
  \href{http://dx.doi.org/10.1103/PhysRevLett.115.141102}{ {\em \prl\/} {\bf
  115} 141102 } [\eprint{1506.09116}]

\bibitem{2016PhRvD..93l4074L}
{Lousto} C~O and {Healy} J 2016
  \href{http://dx.doi.org/10.1103/PhysRevD.93.124074}{ {\em \prd\/} {\bf 93}
  124074 } [\eprint{1601.05086}]

\bibitem{2017CQGra..34f4004G}
{Gerosa} D, {Sperhake} U and {Vo{\v s}mera} J 2017
  \href{http://dx.doi.org/10.1088/1361-6382/aa5e58}{ {\em \cqg\/} {\bf 34}
  064004 } [\eprint{1612.05263}]

\bibitem{2010PhRvD..81h4054K}
{Kesden} M, {Sperhake} U and {Berti} E 2010
  \href{http://dx.doi.org/10.1103/PhysRevD.81.084054}{ {\em \prd\/} {\bf 81}
  084054 } [\eprint{1002.2643}]

\bibitem{2015PhRvD..91d2003V}
{Veitch} J, {Raymond} V, {Farr} B, {Farr} W, {Graff} P, {Vitale} S {\em
  et~al.\/} 2015 \href{http://dx.doi.org/10.1103/PhysRevD.91.042003}{ {\em
  \prd\/} {\bf 91} 042003 } [\eprint{1409.7215}]

\bibitem{2016PhRvX...6d1015A}
{Abbott} B~P {\em et~al.\/} (LIGO Scientific Collaboration, Virgo
  Collaboration) 2016 \href{http://dx.doi.org/10.1103/PhysRevX.6.041015}{ {\em
  \prx\/} {\bf 6} 041015 } [\eprint{1606.04856}]

\bibitem{2016PhRvD..93l4066G}
{Gerosa} D and {Kesden} M 2016
  \href{http://dx.doi.org/10.1103/PhysRevD.93.124066}{ {\em \prd\/} {\bf 93}
  124066 } [\eprint{1605.01067}]

\bibitem{2006PhRvD..74j4005B}
{Buonanno} A, {Chen} Y and {Damour} T 2006
  \href{http://dx.doi.org/10.1103/PhysRevD.74.104005}{ {\em \prd\/} {\bf 74}
  104005 } [\eprint{gr-qc/0508067}]

\bibitem{2009PhRvD..79h4010C}
{Campanelli} M, {Lousto} C~O, {Nakano} H and {Zlochower} Y 2009
  \href{http://dx.doi.org/10.1103/PhysRevD.79.084010}{ {\em \prd\/} {\bf 79}
  084010 } [\eprint{0808.0713}]

\bibitem{2009PhRvD..80h4043B}
{Buonanno} A, {Iyer} B~R, {Ochsner} E, {Pan} Y and {Sathyaprakash} B~S 2009
  \href{http://dx.doi.org/10.1103/PhysRevD.80.084043}{ {\em \prd\/} {\bf 80}
  084043 } [\eprint{0907.0700}]

\bibitem{2015PhRvD..92j4028O}
{Ossokine} S, {Boyle} M, {Kidder} L~E, {Pfeiffer} H~P, {Scheel} M~A and
  {Szil{\'a}gyi} B 2015 \href{http://dx.doi.org/10.1103/PhysRevD.92.104028}{
  {\em \prd\/} {\bf 92} 104028 } [\eprint{1502.01747}]

\bibitem{2016PhRvD..93h4042P}
{P{\"u}rrer} M, {Hannam} M and {Ohme} F 2016
  \href{http://dx.doi.org/10.1103/PhysRevD.93.084042}{ {\em \prd\/} {\bf 93}
  084042 } [\eprint{1512.04955}]

\end{thebibliography}

\end{document}